\documentclass[%
preprint,
superscriptaddress,
amsmath,amssymb,
]{revtex4-1}

\usepackage{graphicx}
\usepackage{dcolumn}
\usepackage{bm,amssymb}
\usepackage{soul,color}

\begin{document}
\renewcommand{\vec}{\mathbf}

\title{Topologically Protected Helical States \\ in Minimally Twisted Bilayer Graphene}

\author{Shengqiang Huang}
\affiliation{Physics Department, University of Arizona, Tucson, AZ 85721, USA}
\author{Kyounghwan Kim}
\affiliation{Microelectronics Research Center, Department of Electrical and Computer Engineering, The University of Texas at Austin, Austin, TX 78758, USA}
\author{Dmitry K. Efimkin}
\affiliation{Department of Physics, The University of Texas at Austin, Austin, TX 78712, USA}
\author{Timothy Lovorn}
\affiliation{Department of Physics, The University of Texas at Austin, Austin, TX 78712, USA}
\author{Takashi Taniguchi}
\affiliation{National Institute for Materials Science, 1- 1 Namiki Tsukuba Ibaraki 305- 0044, Japan}
\author{Kenji Watanabe}
\affiliation{National Institute for Materials Science, 1- 1 Namiki Tsukuba Ibaraki 305- 0044, Japan}
\author{Allan H. MacDonald}
\affiliation{Department of Physics, The University of Texas at Austin, Austin, TX 78712, USA}
\author{Emanuel Tutuc}
\affiliation{Microelectronics Research Center, Department of Electrical and Computer Engineering, The University of Texas at Austin, Austin, TX 78758, USA}
\author{Brian J. LeRoy}
\email{leroy@physics.arizona.edu}
\affiliation{Physics Department, University of Arizona, Tucson, AZ 85721, USA}
\date{\today}

\begin{abstract}
In minimally twisted bilayer graphene, a moir{\'e} pattern consisting of AB and BA stacking regions separated by domain walls forms.  These domain walls are predicted to support counterpropogating topologically protected helical (TPH) edge states when the AB and BA regions are gapped.  We fabricate designer moir{\'e} crystals with wavelengths longer than 50 nm and demonstrate the emergence of TPH states on the domain wall network by scanning tunneling spectroscopy measurements.  We observe a double-line profile of the TPH states on the domain walls, only occurring when the AB and BA regions are gapped. Our results demonstrate a practical and flexible method for TPH state network construction.  
\end{abstract}

\maketitle

\begin{figure}[h]
	\includegraphics[width=8.5cm]{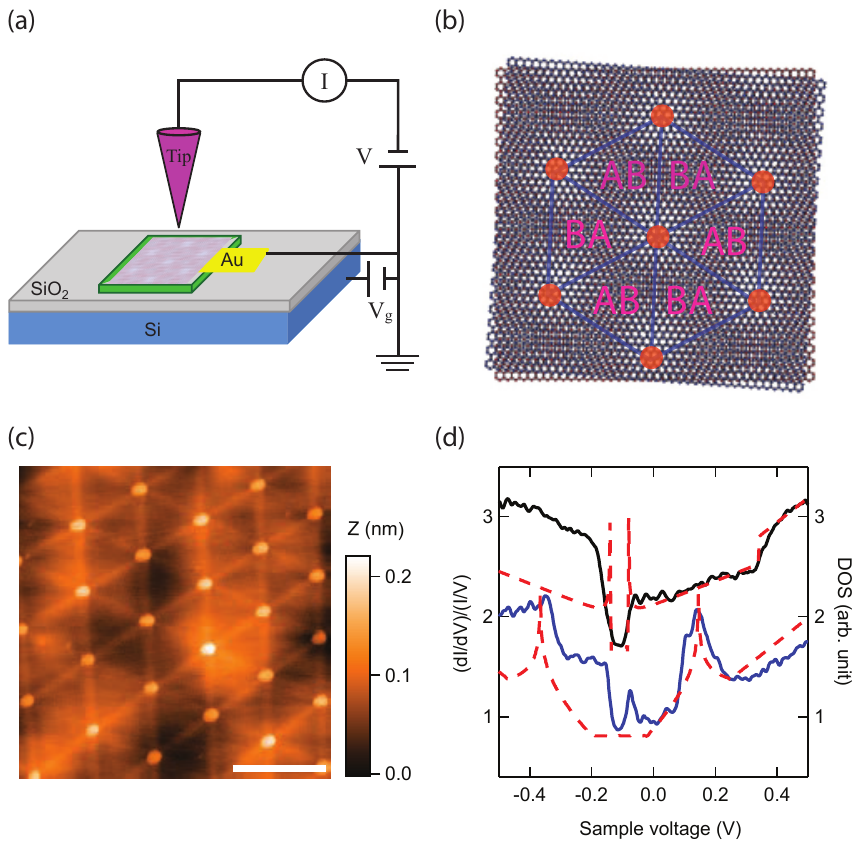} 
	\caption{Measurement schematic, topography and spectroscopy. (a) Schematic diagram of the experimental setup. (b) Local stacking configuration of tBLG. The red spots denote AA points. The blue lines denote domain walls that separate alternating AB and BA regions distinguished by local valley Chern numbers. (c) STM topography image of tBLG with a 57.6 $\pm$ 0.7 nm moir\'e wavelength. The sample voltage is +0.5 V and the tunneling current is 200 pA. The scale bar is 80 nm. (d) $dI/dV/(I/V)$ spectra at the domain wall (solid blue) and at the center of the AB region (solid black). The tip height is stabilized with a current of I = 200 pA at V = +0.5 V.  The back gate voltage is +60 V.  The red dashed lines are the theoretically calculated density of states on the domain wall and in the AB region.  The curves have been vertically offset for clarity.}
	\label{fig:Topography}
\end{figure}

Isolated graphene layers are characterized by low-energy $\pi$-band Dirac cones centered near the K and K' Brillouin-zone corners. Interlayer coupling in bilayers enriches graphene's electronic properties tremendously. For example the bands of Bernal stacked bilayer graphene~\cite{mccann2013electronic}, which has two distinct but equivalent configurations (AB and BA) distinquished by which atom in the top layer unit cell lies directly above an atom in the bottom layer, develop an energy gap and large Berry curvatures when inversion symmetry is broken by an applied transverse electric field~\cite{martin2008topological,yao2009edge,li2010marginality,jung2011valley,zhang2011spontaneous}. Indeed, the total valley Chern number $\mathcal{C}$ of the two occupied valence bands, {\it i.e.} the total Berry curvature integrated over a single valley, is accurately quantized at unit magnitude with a sign that depends on both the sign of the electric field and the stacking configuration~\cite{xiao2010berry,prada2011band,zhang2013valley,san2013helical}.  In uniformly stacked Bernal bilayer graphene, regions with opposite signs of the electric field have opposite valley Chern numbers. Similarly, AB and BA stacked regions of bilayer graphene have opposite Chern numbers under a uniform electric field. In both cases, boundaries between areas with opposite valley Chern numbers support topologically protected helical states~\cite{martin2008topological,qiao2011electronic,wright2011robust,vaezi2013topological,zhang2013valley,san2013helical}. In minimally twisted bilayer graphene (tBLG), a network of domain walls separating AB and BA stacking regions forms~\cite{Hattendorf2013,alden2013strain}. These regions have opposite Chern numbers when inversion symmetry is broken with an external electric field, leading to the prediction of TPH states appearing on the domain walls~\cite{zhang2013valley,san2013helical}. Recently, there have been several efforts to realize TPH states in bilayer graphene devices.  Using split-gates to spatially modulate the electric field, ballistic transport due to TPH states in uniformly stacked bilayer devices has been observed~\cite{li2016gate,li2017valley}. Topologically protected transport along naturally occurring domain walls in Bernal stacked bilayer graphene has also been studied~\cite{ju2015topological,yin2016direct}.  In this letter, we demonstrate experimentally that controllable networks of  
TPH states emerge in minimally twisted bilayer graphene systems. 

tBLG consists of two monolayer graphene layers that are placed on top of each other with a small twist angle applied between their lattices to form a triangular moir\'e pattern~\cite{alden2013strain}.  There has recently been strong experimental interest in electrical transport~\cite{lee2011quantum,sanchez2012quantum,andrei2012electronic,cao2016superlattice,kim2017tunable} and scanning tunneling spectroscopy (STM)~\cite{li2010observation,brihuega2012unraveling,yin2015experimental,wong2015local} studies of these tremendously tunable electronic systems.
The experimental layout for our study of tBLG is sketched in Fig.~\ref{fig:Topography}(a).  The tBLG sample was fabricated through the sequential pick up of two graphene pieces from the same monolayer flake with a controlled rotation angle between them using a thin hexagonal boron nitride (hBN) flake~\cite{kim2016van}. Monolayer graphene and thin ($\sim$10-20 nm) hBN flakes were obtained by exfoliation of bulk flakes onto a SiO$_2$/Si substrate and identified by Raman spectroscopy and atomic force microscopy. A hemispherical handle substrate covered with poly(vinyl alcohol) was employed to first pick up the hBN flake, and then detach consecutively two pieces of monolayer graphene from the same flake. The handle was rotated by a controlled angle before picking up the second graphene piece to create tBLG. The twist angle between the layers is determined by the rotation angle during the pick-up process because the two pieces of monolayer graphene are from the same flake and hence have the same lattice orientation. The whole stacking structure was then placed on a SiO$_2$/Si substrate with the tBLG on top of hBN. The twist angle between the tBLG and hBN was large to ensure that the hBN did not affect the electronic properties of the tBLG. Metal electrodes contacting the tBLG and silicon back gate were written by electron beam lithography followed by deposition of Cr(5 nm)/Au(35 nm) for control of the charge density and interlayer voltage bias. The device was annealed at 350 $^{\circ}{\rm C}$ for 6 hours in high vacuum before putting into the STM which operates under ultrahigh vacuum at a temperature of 4.5 K. Electrochemically etched tungsten tips were used to perform the measurements (see more details about tip calibration and reproducibility in the supplementary information~\cite{SI}).  

The moir\'e pattern wavelength $\lambda$ depends inversely on the twist angle $\theta$ ($\lambda = a/[2 \sin(\theta/2)]$ where $a$ is the graphene lattice constant) and is therefore experimentally controllable.   The moir\'e pattern causes the local stacking configuration of tBLG to vary in space as depicted in Fig.~\ref{fig:Topography}(b). The AA points, marked by red spots, correspond to positions where the hexagons from the two layers lie nearly directly on top of each other. The blue lines connecting AA sites represent domain walls which separate neighboring triangular regions
denoted as AB or BA which have approximate Bernal stacking.

Fig.~\ref{fig:Topography}(c) shows a STM topography image of a tBLG sample with a $\lambda=57.6 \pm 0.7\; \hbox{nm}$ moir\'e wavelength, which corresponds to a twist angle of 0.245 $\pm$ 0.003 degrees. The AA sites appear as bright spots, in agreement with previous observations~\cite{li2010observation,brihuega2012unraveling,wong2015local,yin2015experimental,kim2017tunable} (see more details about the AA region in the supplementary information~\cite{SI}). The visible network of lines connecting AA sites are domain walls of several nanometer width that can host topologically protected modes as we will show below.  The topography varies with sample voltage indicating that the bright line features in the topography are due to density of states (DOS) variations rather than physical topography variations in the sample (see more details in the supplementary information~\cite{SI}). 

To examine the variation of the density of states with position in tBLG, we have performed spectroscopy measurements at various locations within the moir\'e unit cell.  This is done by fixing the location of the tip and then turning off the feedback circuit so that the tip height remains constant.  A small ac voltage (3 mV, 572 Hz) is added to the sample voltage and the differential conductance, $dI/dV$, is measured as a function of sample voltage.  Fig.~\ref{fig:Topography}(d) shows the normalized differential conductance, $dI/dV/(I/V)$, spectra at the center of an AB region (black) and on a domain wall (blue) with a back gate voltage of 60 V. The spectra for AB and BA regions are identical (see more details in the supplementary information~\cite{SI}). The STS spectrum in the AB region shows a nearly constant density of states with a dip around -0.1 V which corresponds to the band gap which is opened by the electric field.  There is also an increase around 0.3 V which corresponds to the energy of the second subband in the conduction band.  The spectrum on the domain wall is qualitatively different.  Instead of the pronounced dip around -0.1 V corresponding to the band gap, there are two peaks symmetrically found around $\pm$ 0.25 V from the charge neutrality point.

The measured STS spectrum has interesting features both at energies below the band gap $E\ll\Delta$ where $\Delta$ is the band gap in the AB (BA) regions~\cite{SI} and at high energy $E\gtrsim \Delta$ that have different origin.  At $E\ll\Delta$ the only relevant degrees of freedom are helical states along the domain walls. Their interference results in the domain wall pattern band structure that has been theoretically described by network models~\cite{EfimkinNetwork,san2013helical}.  The theories~\cite{EfimkinNetwork,san2013helical} predict the presence of a set of features in the density of states with characteristic energy scale $E_\mathrm{\lambda}=2\pi \hbar v/3 \lambda \approx 24 \; \hbox{meV}$ where $v$ is the velocity of Dirac fermions. In our experiment the condition $E_\mathrm{\lambda}\ll \Delta$ is not so well satisfied and only the peak at -0.05 V in Fig.1~(d) is well resolved and can be attributed to these predictions. To explain the high energy features in the STS spectrum for long wavelength moir\'e patterns we use the local electronic picture of the continuum model~\cite{bistritzer2011moire} for twisted bilayer graphene that is valid irrespective of whether or not the bilayer structure is commensurate at the atomic level. Its Hamiltonian,
\begin{equation*}
\label{HamiltonianFourBands}
H=
\begin{pmatrix}
	v \sigma_\mathrm{t} \vec{p}-u & T(\vec{r})
	\\ T^+(\vec{r}) & v \sigma_\mathrm{b} \vec{p}+u
\end{pmatrix},
\end{equation*}
acts in the sublattice space ($\mathrm{A}$ and $\mathrm{B}$) of the system  $\psi=\{\psi_\mathrm{A}^\mathrm{t},\psi_\mathrm{B}^\mathrm{t}, \psi_\mathrm{A}^\mathrm{b},\psi_{\mathrm{B}}^\mathrm{b}\}$, where $\mathrm{t}$ ($\mathrm{b}$) refers to the top (bottom) layer; $\sigma_\mathrm{t(b)}$ is the vector of Pauli matrices rotated by the angle $\theta/2$ ($-\theta/2$); and $2 u$ is the energy splitting of Dirac points induced by an electric field perpendicular to the bilayer. $T(\vec{r})$ is the tunneling matrix accounting for hybridization between layers, which is periodic with the period of the moir$\mathrm{\acute{e}}$ pattern (please see details in the supplementary information~\cite{SI}). Valley and spin degrees of freedom are decoupled and the electronic spectrum therefore has fourfold degeneracy. 


In an electric field, the LDOS is gapless only along the domain wall lines.  The gapped local bands have valley Chern numbers that are quantized throughout and have opposite signs in AB and BA regions leading to the formation of topologically protected states along their interface. The calculated local density of states at the domain wall and in the AB/BA region are shown by the dashed lines in Fig.~\ref{fig:Topography}(d). The theoretical model captures all the principal observed features at energies $E\gtrsim \Delta$ in the local density of states. The spectrum of the AB/BA region has a gap under a bias. The spectrum of the domain wall is gapless and has two saddle points at $E\approx\pm 250\;\hbox{meV}$, which are responsible for the observed van Hove singularities of the electronic density of states clearly seen in the experimental data. 

\begin{figure}[h]
	\includegraphics[width=8.5cm]{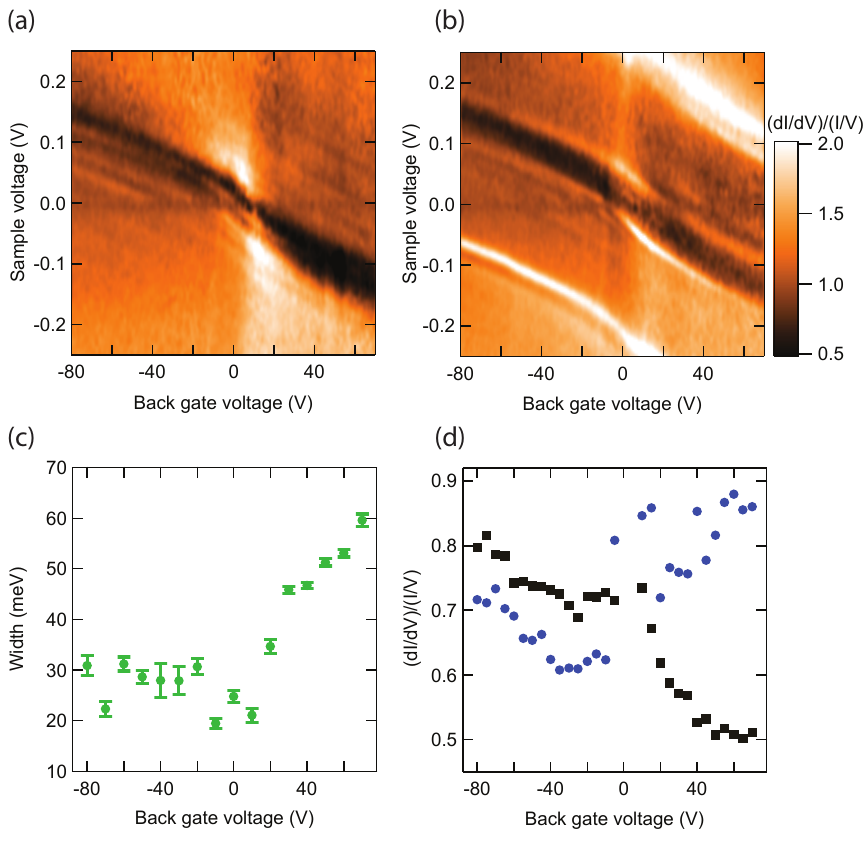} 
	\caption{Gate voltage dependent spectroscopy. (a)-(b) $dI/dV/(I/V)$ spectra as a function of sample voltage and back gate voltage at the center of the AB region (a) and the domain wall (b) respectively. The tip height is stabilized with a current of I = 200 pA at V = +0.5 V. (c) Band gap width in the AB region as a function of back gate voltage. The error bars represent the uncertainty of the gap width when fitting the $dI/dV/(I/V)$ spectra. (d) Minimum value of the $dI/dV/(I/V)$ spectra in the AB region (black) and the domain wall (blue).}
	\label{fig:Gatesweep}
\end{figure}

We have performed spectroscopy measurements as a function of back gate voltage to investigate the dependence of the density of states on the inversion-symmetry breaking electric field. The electric field has contributions from the back gate voltage, the tip bias and the work function mismatch between the tip and tBLG.  The latter two give rise to an offset in gate voltage of the zero in electric field.  As expected, a sizable band gap is opened in the AB region as the back gate voltage and hence the electric field is increased as shown in Fig.~\ref{fig:Gatesweep}(a).  Fig.~\ref{fig:Gatesweep}(c) shows the measured band gap in the AB region as a function of back gate voltage (see supplementary information~\cite{SI} for details on the extraction of the band gap).  The band gap is largest at large positive back gate voltages, while it diminishes as the back gate voltage decreases.  In contrast the $dI/dV/(I/V)$ spectra on the domain wall responds weakly to a varying electric field as shown in Fig.~\ref{fig:Gatesweep}(b). Fig.~\ref{fig:Gatesweep}(d) shows the minimum value of the $dI/dV/(I/V)$ spectra as a function of the back gate voltage in the AB region (black) and the domain wall (blue) respectively.  The minimum in the AB region remains constant for negative back gate voltages but decreases as the back gate voltage becomes positive and the band gap opens. While the minimum on the domain wall increases with the back gate voltage, consistent with a DOS that is enhanced due to the appearance of TPH states on the domain wall as the band gap is opened.

\begin{figure}[h]
	\includegraphics[width=14cm]{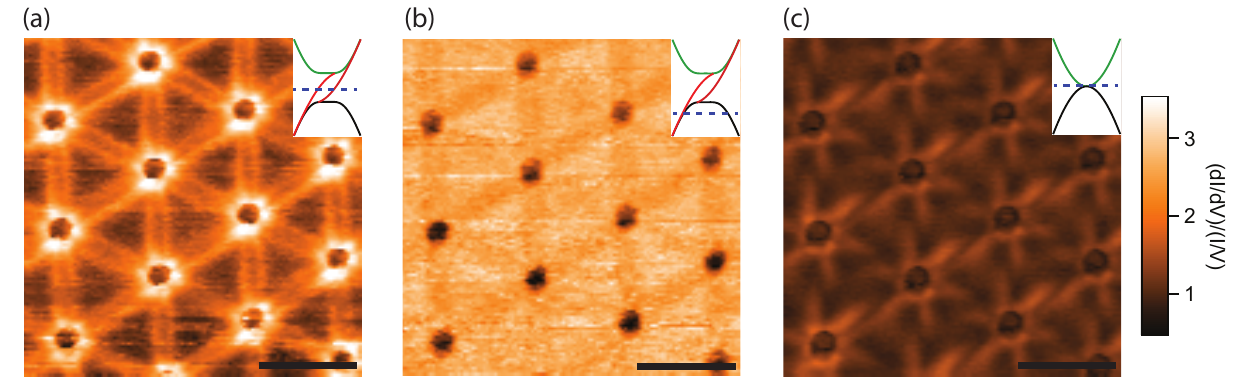} 
	\caption{Spatially resolved density of states. (a)-(c) Density of states mapping with (a) sample voltage -0.11 V and back gate voltage +60 V, (b) sample voltage -0.245V and back gate voltage +60 V, and (c) sample voltage 0.1 V and back gate voltage -50 V. The insets are schematics of the band structure, with the sample voltage being probed marked by the blue dashed line. The green and black curves denote the conduction and valence bands of the Bernal stacked bilayer regions respectively while the red lines are due to the TPH states on the domain wall.  In all figures, the tunnel current is 200 pA, the ac voltage is 8 mV and the scale bar is 50 nm.}
	\label{fig:DOSmapping}
\end{figure}

To further confirm the presence of TPH states on the domain walls, we perform local density of states (LDOS) mapping measurements at specific energies to image and compare the states on the domain walls and in the Bernal stacked regions. Fig.~\ref{fig:DOSmapping}(a) is a LDOS map acquired at the energy of the center of the band gap with a back gate voltage of +60 V.  A double line network of enhanced density of states connecting AA regions is clearly visible, while the AB and BA regions are much darker due to the band gap in these regions. This enhanced DOS along domain walls separating AB and BA stacked regions demonstrates the presence of TPH states as had been previously theoretically predicted~\cite{san2013helical}. Their spatial separation manifesting in the double-line profile has been reported previously for isolated domain walls and agrees with the predictions for helical states~\cite{yin2016direct}. The inset of Fig.~\ref{fig:DOSmapping}(a) is a schematic showing the band structure in one valley.  The green and black curves denote the conduction and valence bands of the Bernal stacked bilayer graphene while the red curves are due to the TPH states on the domain wall.  The sample voltage is in the middle of the band gap as indicated by the dashed blue line.  When the energy is changed to be outside the band gap of the Bernal stacked region as shown in the inset of Fig.~\ref{fig:DOSmapping}(b), the contrast between the domain wall and the AB or BA region is much smaller.  This indicates that the TPH arise at energies where the AB and BA regions are gapped and outside this energy the states are delocalized away from the AA sites.  Furthermore, at a back gate voltage of -50 V where the electric field is negligibly small, the contrast between the domain wall and the AB and BA regions is still weaker as shown in Fig.~\ref{fig:DOSmapping}(c).  Once again, this demonstrates that TPH states do not occur when the AB and BA regions are gapless, as seen in the inset of Fig.~\ref{fig:DOSmapping}(c).  This is once again in agreement with the theoretical prediction shown in Fig. 2(f) of the work by San-Jose and Prada of a depleted area around the AA sites and an intricate pattern surrounding them~\cite{san2013helical}.
 
\begin{figure}[h]
	\includegraphics[width=8.5cm]{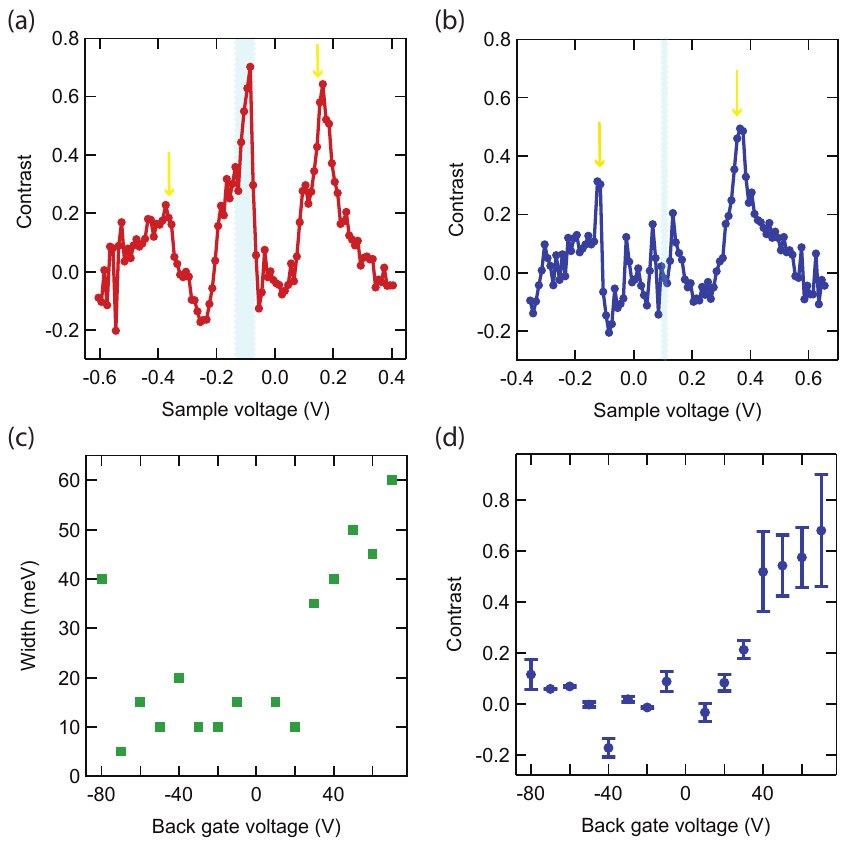} 
	\caption{Strength of TPH states. (a)-(b) Contrast between the domain wall and the AB region as a function of sample voltage with back gate voltage (a) +60 V and (b) -50 V. The blue shaded regions represent the band gap and charge neutrality point respectively.  The orange arrows point towards the increased DOS at high energy on the domain walls.  (c) Full width at half maximum of the central peak in contrast as a function of the back gate voltage. (d) Averaged contrast over the FWHM as a function of the back gate voltage. The error bars represent the standard deviation of averaged values.}
	\label{fig:contrast}
\end{figure}

In order to quantify the strength of the TPH states, we define the contrast as $C = \frac{I(DW)- I(AB)}{I(AB)}$, where $I(DW)$ and $I(AB)$ are the intensity of the DOS on the domain wall and AB region respectively. Fig.~\ref{fig:contrast}(a) shows the contrast as a function of sample voltage at a back gate voltage of +60 V. The central peak corresponds to the band gap in the AB region where the contrast is enhanced due to the presence of the TPH states. The contrast is strongest in the center of the gap and drops to near zero at the conduction and valence band edges in the AB region (see supplementary information~\cite{SI} for the determination of the conduction band edge using Fourier transform scanning tunneling spectroscopy). There are two additional peaks at higher and lower voltages that are due to the enhanced DOS of the domain wall that occurs at $\pm$ 250 meV from the charge neutrality point as seen in Fig.~\ref{fig:Gatesweep}(b) (see supplementary information~\cite{SI} for discussion of these peaks).    The central peak disappears at a back gate voltage of -50 V as seen in Fig.~\ref{fig:contrast}(b) since there is no electric field and no TPH states occur, but the two additional peaks are still visible. Fig.~\ref{fig:contrast}(c) shows the full width at half maximum of the central peak in the contrast which decreases with the decreasing back gate voltage. Over this energy range, the averaged contrast value as plotted in Fig.~\ref{fig:contrast}(d) was taken to represent the strength of the TPH states. The value drops as the back gate voltage decreases and becomes very small at negative back gate voltages where the band gap is not well developed. These results demonstrate that the TPH states only occur when a band gap is opened in the Bernal stacked regions and their strength is correlated with the size of this gap as show in Fig.~\ref{fig:Gatesweep}(c). 

In summary, we realized tBLG samples with moir\'e wavelengths longer than 50 nm, long enough to form well defined topologically distinct regions within the sample when an inversion-symmetry breaking electric field is applied.  We observe TPH states along boundaries between these regions because they have non-zero density-of-states within the topological states gaps, and also characteristic features in the LDOS at higher energies.  The strength of the TPH states and the energy range over which they appear correlate with the electric field and band gap in the AB and BA regions. The network of domain walls in tBLG can be used as topological channels to transfer charge in a dissipationless manner. The length of the domain walls depends on the twist angle between the two graphene layers which can be controlled during fabrication, enabling controllable topological networks for future applications.    \\

\begin{acknowledgments}

The work at the University of Arizona was partially supported by the U. S. Army Research Laboratory and the U. S. Army Research Office under contract/grant number W911NF-14-1-0653 and the National Science Foundation under grant EECS-1607911.  The work at the University of Texas was  supported by the National Science Foundation under grant EECS-1610008, and by the  Army Research Office under Award W911NF-17-1-0312.  K.W. and T.T. acknowledge support from the Elemental Strategy Initiative conducted by the MEXT, Japan and JSPS KAKENHI Grant Numbers JP15K21722.

\nocite{trambly2010localization,yankowitz2014band,castro2007biased,kuzmenko2009infrared,li2009band,zhang2008determination,blochl1994tetrahedron}

\end{acknowledgments}

\bibliography{ReferenceTPStBLG}

\begin{thebibliography}{39}%
\makeatletter
\providecommand \@ifxundefined [1]{%
 \@ifx{#1\undefined}
}%
\providecommand \@ifnum [1]{%
 \ifnum #1\expandafter \@firstoftwo
 \else \expandafter \@secondoftwo
 \fi
}%
\providecommand \@ifx [1]{%
 \ifx #1\expandafter \@firstoftwo
 \else \expandafter \@secondoftwo
 \fi
}%
\providecommand \natexlab [1]{#1}%
\providecommand \enquote  [1]{``#1''}%
\providecommand \bibnamefont  [1]{#1}%
\providecommand \bibfnamefont [1]{#1}%
\providecommand \citenamefont [1]{#1}%
\providecommand \href@noop [0]{\@secondoftwo}%
\providecommand \href [0]{\begingroup \@sanitize@url \@href}%
\providecommand \@href[1]{\@@startlink{#1}\@@href}%
\providecommand \@@href[1]{\endgroup#1\@@endlink}%
\providecommand \@sanitize@url [0]{\catcode `\\12\catcode `\$12\catcode
  `\&12\catcode `\#12\catcode `\^12\catcode `\_12\catcode `\%12\relax}%
\providecommand \@@startlink[1]{}%
\providecommand \@@endlink[0]{}%
\providecommand \url  [0]{\begingroup\@sanitize@url \@url }%
\providecommand \@url [1]{\endgroup\@href {#1}{\urlprefix }}%
\providecommand \urlprefix  [0]{URL }%
\providecommand \Eprint [0]{\href }%
\providecommand \doibase [0]{http://dx.doi.org/}%
\providecommand \selectlanguage [0]{\@gobble}%
\providecommand \bibinfo  [0]{\@secondoftwo}%
\providecommand \bibfield  [0]{\@secondoftwo}%
\providecommand \translation [1]{[#1]}%
\providecommand \BibitemOpen [0]{}%
\providecommand \bibitemStop [0]{}%
\providecommand \bibitemNoStop [0]{.\EOS\space}%
\providecommand \EOS [0]{\spacefactor3000\relax}%
\providecommand \BibitemShut  [1]{\csname bibitem#1\endcsname}%
\let\auto@bib@innerbib\@empty
\bibitem [{\citenamefont {McCann}\ and\ \citenamefont
  {Koshino}(2013)}]{mccann2013electronic}%
  \BibitemOpen
  \bibfield  {author} {\bibinfo {author} {\bibfnamefont {E.}~\bibnamefont
  {McCann}}\ and\ \bibinfo {author} {\bibfnamefont {M.}~\bibnamefont
  {Koshino}},\ }\href@noop {} {\bibfield  {journal} {\bibinfo  {journal}
  {Reports on Progress in Physics}\ }\textbf {\bibinfo {volume} {76}},\
  \bibinfo {pages} {056503} (\bibinfo {year} {2013})}\BibitemShut {NoStop}%
\bibitem [{\citenamefont {Martin}\ \emph {et~al.}(2008)\citenamefont {Martin},
  \citenamefont {Blanter},\ and\ \citenamefont
  {Morpurgo}}]{martin2008topological}%
  \BibitemOpen
  \bibfield  {author} {\bibinfo {author} {\bibfnamefont {I.}~\bibnamefont
  {Martin}}, \bibinfo {author} {\bibfnamefont {Y.~M.}\ \bibnamefont {Blanter}},
  \ and\ \bibinfo {author} {\bibfnamefont {A.~F.}\ \bibnamefont {Morpurgo}},\
  }\href@noop {} {\bibfield  {journal} {\bibinfo  {journal} {Physical Review
  Letters}\ }\textbf {\bibinfo {volume} {100}},\ \bibinfo {pages} {036804}
  (\bibinfo {year} {2008})}\BibitemShut {NoStop}%
\bibitem [{\citenamefont {Yao}\ \emph {et~al.}(2009)\citenamefont {Yao},
  \citenamefont {Yang},\ and\ \citenamefont {Niu}}]{yao2009edge}%
  \BibitemOpen
  \bibfield  {author} {\bibinfo {author} {\bibfnamefont {W.}~\bibnamefont
  {Yao}}, \bibinfo {author} {\bibfnamefont {S.~A.}\ \bibnamefont {Yang}}, \
  and\ \bibinfo {author} {\bibfnamefont {Q.}~\bibnamefont {Niu}},\ }\href@noop
  {} {\bibfield  {journal} {\bibinfo  {journal} {Physical Review Letters}\
  }\textbf {\bibinfo {volume} {102}},\ \bibinfo {pages} {096801} (\bibinfo
  {year} {2009})}\BibitemShut {NoStop}%
\bibitem [{\citenamefont {Li}\ \emph {et~al.}(2010{\natexlab{a}})\citenamefont
  {Li}, \citenamefont {Morpurgo}, \citenamefont {B{\"u}ttiker},\ and\
  \citenamefont {Martin}}]{li2010marginality}%
  \BibitemOpen
  \bibfield  {author} {\bibinfo {author} {\bibfnamefont {J.}~\bibnamefont
  {Li}}, \bibinfo {author} {\bibfnamefont {A.~F.}\ \bibnamefont {Morpurgo}},
  \bibinfo {author} {\bibfnamefont {M.}~\bibnamefont {B{\"u}ttiker}}, \ and\
  \bibinfo {author} {\bibfnamefont {I.}~\bibnamefont {Martin}},\ }\href@noop {}
  {\bibfield  {journal} {\bibinfo  {journal} {Physical Review B}\ }\textbf
  {\bibinfo {volume} {82}},\ \bibinfo {pages} {245404} (\bibinfo {year}
  {2010}{\natexlab{a}})}\BibitemShut {NoStop}%
\bibitem [{\citenamefont {Jung}\ \emph {et~al.}(2011)\citenamefont {Jung},
  \citenamefont {Zhang}, \citenamefont {Qiao},\ and\ \citenamefont
  {MacDonald}}]{jung2011valley}%
  \BibitemOpen
  \bibfield  {author} {\bibinfo {author} {\bibfnamefont {J.}~\bibnamefont
  {Jung}}, \bibinfo {author} {\bibfnamefont {F.}~\bibnamefont {Zhang}},
  \bibinfo {author} {\bibfnamefont {Z.}~\bibnamefont {Qiao}}, \ and\ \bibinfo
  {author} {\bibfnamefont {A.~H.}\ \bibnamefont {MacDonald}},\ }\href@noop {}
  {\bibfield  {journal} {\bibinfo  {journal} {Physical Review B}\ }\textbf
  {\bibinfo {volume} {84}},\ \bibinfo {pages} {075418} (\bibinfo {year}
  {2011})}\BibitemShut {NoStop}%
\bibitem [{\citenamefont {Zhang}\ \emph {et~al.}(2011)\citenamefont {Zhang},
  \citenamefont {Jung}, \citenamefont {Fiete}, \citenamefont {Niu},\ and\
  \citenamefont {MacDonald}}]{zhang2011spontaneous}%
  \BibitemOpen
  \bibfield  {author} {\bibinfo {author} {\bibfnamefont {F.}~\bibnamefont
  {Zhang}}, \bibinfo {author} {\bibfnamefont {J.}~\bibnamefont {Jung}},
  \bibinfo {author} {\bibfnamefont {G.~A.}\ \bibnamefont {Fiete}}, \bibinfo
  {author} {\bibfnamefont {Q.}~\bibnamefont {Niu}}, \ and\ \bibinfo {author}
  {\bibfnamefont {A.~H.}\ \bibnamefont {MacDonald}},\ }\href@noop {} {\bibfield
   {journal} {\bibinfo  {journal} {Physical Review Letters}\ }\textbf {\bibinfo
  {volume} {106}},\ \bibinfo {pages} {156801} (\bibinfo {year}
  {2011})}\BibitemShut {NoStop}%
\bibitem [{\citenamefont {Xiao}\ \emph {et~al.}(2010)\citenamefont {Xiao},
  \citenamefont {Chang},\ and\ \citenamefont {Niu}}]{xiao2010berry}%
  \BibitemOpen
  \bibfield  {author} {\bibinfo {author} {\bibfnamefont {D.}~\bibnamefont
  {Xiao}}, \bibinfo {author} {\bibfnamefont {M.-C.}\ \bibnamefont {Chang}}, \
  and\ \bibinfo {author} {\bibfnamefont {Q.}~\bibnamefont {Niu}},\ }\href@noop
  {} {\bibfield  {journal} {\bibinfo  {journal} {Reviews of Modern Physics}\
  }\textbf {\bibinfo {volume} {82}},\ \bibinfo {pages} {1959} (\bibinfo {year}
  {2010})}\BibitemShut {NoStop}%
\bibitem [{\citenamefont {Prada}\ \emph {et~al.}(2011)\citenamefont {Prada},
  \citenamefont {San-Jose}, \citenamefont {Brey},\ and\ \citenamefont
  {Fertig}}]{prada2011band}%
  \BibitemOpen
  \bibfield  {author} {\bibinfo {author} {\bibfnamefont {E.}~\bibnamefont
  {Prada}}, \bibinfo {author} {\bibfnamefont {P.}~\bibnamefont {San-Jose}},
  \bibinfo {author} {\bibfnamefont {L.}~\bibnamefont {Brey}}, \ and\ \bibinfo
  {author} {\bibfnamefont {H.}~\bibnamefont {Fertig}},\ }\href@noop {}
  {\bibfield  {journal} {\bibinfo  {journal} {Solid State Communications}\
  }\textbf {\bibinfo {volume} {151}},\ \bibinfo {pages} {1075} (\bibinfo {year}
  {2011})}\BibitemShut {NoStop}%
\bibitem [{\citenamefont {Zhang}\ \emph {et~al.}(2013)\citenamefont {Zhang},
  \citenamefont {MacDonald},\ and\ \citenamefont {Mele}}]{zhang2013valley}%
  \BibitemOpen
  \bibfield  {author} {\bibinfo {author} {\bibfnamefont {F.}~\bibnamefont
  {Zhang}}, \bibinfo {author} {\bibfnamefont {A.~H.}\ \bibnamefont
  {MacDonald}}, \ and\ \bibinfo {author} {\bibfnamefont {E.~J.}\ \bibnamefont
  {Mele}},\ }\href@noop {} {\bibfield  {journal} {\bibinfo  {journal}
  {Proceedings of the National Academy of Sciences USA}\ }\textbf {\bibinfo
  {volume} {110}},\ \bibinfo {pages} {10546} (\bibinfo {year}
  {2013})}\BibitemShut {NoStop}%
\bibitem [{\citenamefont {San-Jose}\ and\ \citenamefont
  {Prada}(2013)}]{san2013helical}%
  \BibitemOpen
  \bibfield  {author} {\bibinfo {author} {\bibfnamefont {P.}~\bibnamefont
  {San-Jose}}\ and\ \bibinfo {author} {\bibfnamefont {E.}~\bibnamefont
  {Prada}},\ }\href@noop {} {\bibfield  {journal} {\bibinfo  {journal}
  {Physical Review B}\ }\textbf {\bibinfo {volume} {88}},\ \bibinfo {pages}
  {121408} (\bibinfo {year} {2013})}\BibitemShut {NoStop}%
\bibitem [{\citenamefont {Qiao}\ \emph {et~al.}(2011)\citenamefont {Qiao},
  \citenamefont {Jung}, \citenamefont {Niu},\ and\ \citenamefont
  {MacDonald}}]{qiao2011electronic}%
  \BibitemOpen
  \bibfield  {author} {\bibinfo {author} {\bibfnamefont {Z.}~\bibnamefont
  {Qiao}}, \bibinfo {author} {\bibfnamefont {J.}~\bibnamefont {Jung}}, \bibinfo
  {author} {\bibfnamefont {Q.}~\bibnamefont {Niu}}, \ and\ \bibinfo {author}
  {\bibfnamefont {A.~H.}\ \bibnamefont {MacDonald}},\ }\href@noop {} {\bibfield
   {journal} {\bibinfo  {journal} {Nano Letters}\ }\textbf {\bibinfo {volume}
  {11}},\ \bibinfo {pages} {3453} (\bibinfo {year} {2011})}\BibitemShut
  {NoStop}%
\bibitem [{\citenamefont {Wright}\ and\ \citenamefont
  {Hyart}(2011)}]{wright2011robust}%
  \BibitemOpen
  \bibfield  {author} {\bibinfo {author} {\bibfnamefont {A.~R.}\ \bibnamefont
  {Wright}}\ and\ \bibinfo {author} {\bibfnamefont {T.}~\bibnamefont {Hyart}},\
  }\href@noop {} {\bibfield  {journal} {\bibinfo  {journal} {Applied Physics
  Letters}\ }\textbf {\bibinfo {volume} {98}},\ \bibinfo {pages} {251902}
  (\bibinfo {year} {2011})}\BibitemShut {NoStop}%
\bibitem [{\citenamefont {Vaezi}\ \emph {et~al.}(2013)\citenamefont {Vaezi},
  \citenamefont {Liang}, \citenamefont {Ngai}, \citenamefont {Yang},\ and\
  \citenamefont {Kim}}]{vaezi2013topological}%
  \BibitemOpen
  \bibfield  {author} {\bibinfo {author} {\bibfnamefont {A.}~\bibnamefont
  {Vaezi}}, \bibinfo {author} {\bibfnamefont {Y.}~\bibnamefont {Liang}},
  \bibinfo {author} {\bibfnamefont {D.~H.}\ \bibnamefont {Ngai}}, \bibinfo
  {author} {\bibfnamefont {L.}~\bibnamefont {Yang}}, \ and\ \bibinfo {author}
  {\bibfnamefont {E.-A.}\ \bibnamefont {Kim}},\ }\href@noop {} {\bibfield
  {journal} {\bibinfo  {journal} {Physical Review X}\ }\textbf {\bibinfo
  {volume} {3}},\ \bibinfo {pages} {021018} (\bibinfo {year}
  {2013})}\BibitemShut {NoStop}%
\bibitem [{\citenamefont {Hattendorf}\ \emph {et~al.}(2013)\citenamefont
  {Hattendorf}, \citenamefont {Georgi}, \citenamefont {Liebmann},\ and\
  \citenamefont {Morgenstern}}]{Hattendorf2013}%
  \BibitemOpen
  \bibfield  {author} {\bibinfo {author} {\bibfnamefont {S.}~\bibnamefont
  {Hattendorf}}, \bibinfo {author} {\bibfnamefont {A.}~\bibnamefont {Georgi}},
  \bibinfo {author} {\bibfnamefont {M.}~\bibnamefont {Liebmann}}, \ and\
  \bibinfo {author} {\bibfnamefont {M.}~\bibnamefont {Morgenstern}},\
  }\href@noop {} {\bibfield  {journal} {\bibinfo  {journal} {Surface Science}\
  }\textbf {\bibinfo {volume} {610}},\ \bibinfo {pages} {53} (\bibinfo {year}
  {2013})}\BibitemShut {NoStop}%
\bibitem [{\citenamefont {Alden}\ \emph {et~al.}(2013)\citenamefont {Alden},
  \citenamefont {Tsen}, \citenamefont {Huang}, \citenamefont {Hovden},
  \citenamefont {Brown}, \citenamefont {Park}, \citenamefont {Muller},\ and\
  \citenamefont {McEuen}}]{alden2013strain}%
  \BibitemOpen
  \bibfield  {author} {\bibinfo {author} {\bibfnamefont {J.~S.}\ \bibnamefont
  {Alden}}, \bibinfo {author} {\bibfnamefont {A.~W.}\ \bibnamefont {Tsen}},
  \bibinfo {author} {\bibfnamefont {P.~Y.}\ \bibnamefont {Huang}}, \bibinfo
  {author} {\bibfnamefont {R.}~\bibnamefont {Hovden}}, \bibinfo {author}
  {\bibfnamefont {L.}~\bibnamefont {Brown}}, \bibinfo {author} {\bibfnamefont
  {J.}~\bibnamefont {Park}}, \bibinfo {author} {\bibfnamefont {D.~A.}\
  \bibnamefont {Muller}}, \ and\ \bibinfo {author} {\bibfnamefont {P.~L.}\
  \bibnamefont {McEuen}},\ }\href@noop {} {\bibfield  {journal} {\bibinfo
  {journal} {Proceedings of the National Academy of Sciences USA}\ }\textbf
  {\bibinfo {volume} {110}},\ \bibinfo {pages} {11256} (\bibinfo {year}
  {2013})}\BibitemShut {NoStop}%
\bibitem [{\citenamefont {Li}\ \emph {et~al.}(2016)\citenamefont {Li},
  \citenamefont {Wang}, \citenamefont {McFaul}, \citenamefont {Zern},
  \citenamefont {Ren}, \citenamefont {Watanabe}, \citenamefont {Taniguchi},
  \citenamefont {Qiao},\ and\ \citenamefont {Zhu}}]{li2016gate}%
  \BibitemOpen
  \bibfield  {author} {\bibinfo {author} {\bibfnamefont {J.}~\bibnamefont
  {Li}}, \bibinfo {author} {\bibfnamefont {K.}~\bibnamefont {Wang}}, \bibinfo
  {author} {\bibfnamefont {K.~J.}\ \bibnamefont {McFaul}}, \bibinfo {author}
  {\bibfnamefont {Z.}~\bibnamefont {Zern}}, \bibinfo {author} {\bibfnamefont
  {Y.}~\bibnamefont {Ren}}, \bibinfo {author} {\bibfnamefont {K.}~\bibnamefont
  {Watanabe}}, \bibinfo {author} {\bibfnamefont {T.}~\bibnamefont {Taniguchi}},
  \bibinfo {author} {\bibfnamefont {Z.}~\bibnamefont {Qiao}}, \ and\ \bibinfo
  {author} {\bibfnamefont {J.}~\bibnamefont {Zhu}},\ }\href@noop {} {\bibfield
  {journal} {\bibinfo  {journal} {Nature Nanotechnology}\ }\textbf {\bibinfo
  {volume} {11}},\ \bibinfo {pages} {1060} (\bibinfo {year}
  {2016})}\BibitemShut {NoStop}%
\bibitem [{\citenamefont {Li}\ \emph {et~al.}(2017)\citenamefont {Li},
  \citenamefont {Zhang}, \citenamefont {Yin}, \citenamefont {Zhang},
  \citenamefont {Watanabe}, \citenamefont {Taniguchi}, \citenamefont {Liu},\
  and\ \citenamefont {Zhu}}]{li2017valley}%
  \BibitemOpen
  \bibfield  {author} {\bibinfo {author} {\bibfnamefont {J.}~\bibnamefont
  {Li}}, \bibinfo {author} {\bibfnamefont {R.-X.}\ \bibnamefont {Zhang}},
  \bibinfo {author} {\bibfnamefont {Z.}~\bibnamefont {Yin}}, \bibinfo {author}
  {\bibfnamefont {J.}~\bibnamefont {Zhang}}, \bibinfo {author} {\bibfnamefont
  {K.}~\bibnamefont {Watanabe}}, \bibinfo {author} {\bibfnamefont
  {T.}~\bibnamefont {Taniguchi}}, \bibinfo {author} {\bibfnamefont
  {C.}~\bibnamefont {Liu}}, \ and\ \bibinfo {author} {\bibfnamefont
  {J.}~\bibnamefont {Zhu}},\ }\href@noop {} {\bibfield  {journal} {\bibinfo
  {journal} {arXiv preprint arXiv:1708.02311}\ } (\bibinfo {year}
  {2017})}\BibitemShut {NoStop}%
\bibitem [{\citenamefont {Ju}\ \emph {et~al.}(2015)\citenamefont {Ju},
  \citenamefont {Shi}, \citenamefont {Nair}, \citenamefont {Lv}, \citenamefont
  {Jin}, \citenamefont {Velasco~Jr}, \citenamefont {Ojeda-Aristizabal},
  \citenamefont {Bechtel}, \citenamefont {Martin}, \citenamefont {Zettl} \emph
  {et~al.}}]{ju2015topological}%
  \BibitemOpen
  \bibfield  {author} {\bibinfo {author} {\bibfnamefont {L.}~\bibnamefont
  {Ju}}, \bibinfo {author} {\bibfnamefont {Z.}~\bibnamefont {Shi}}, \bibinfo
  {author} {\bibfnamefont {N.}~\bibnamefont {Nair}}, \bibinfo {author}
  {\bibfnamefont {Y.}~\bibnamefont {Lv}}, \bibinfo {author} {\bibfnamefont
  {C.}~\bibnamefont {Jin}}, \bibinfo {author} {\bibfnamefont {J.}~\bibnamefont
  {Velasco~Jr}}, \bibinfo {author} {\bibfnamefont {C.}~\bibnamefont
  {Ojeda-Aristizabal}}, \bibinfo {author} {\bibfnamefont {H.~A.}\ \bibnamefont
  {Bechtel}}, \bibinfo {author} {\bibfnamefont {M.~C.}\ \bibnamefont {Martin}},
  \bibinfo {author} {\bibfnamefont {A.}~\bibnamefont {Zettl}},  \emph
  {et~al.},\ }\href@noop {} {\bibfield  {journal} {\bibinfo  {journal}
  {Nature}\ }\textbf {\bibinfo {volume} {520}},\ \bibinfo {pages} {650}
  (\bibinfo {year} {2015})}\BibitemShut {NoStop}%
\bibitem [{\citenamefont {Yin}\ \emph {et~al.}(2016)\citenamefont {Yin},
  \citenamefont {Jiang}, \citenamefont {Qiao},\ and\ \citenamefont
  {He}}]{yin2016direct}%
  \BibitemOpen
  \bibfield  {author} {\bibinfo {author} {\bibfnamefont {L.-J.}\ \bibnamefont
  {Yin}}, \bibinfo {author} {\bibfnamefont {H.}~\bibnamefont {Jiang}}, \bibinfo
  {author} {\bibfnamefont {J.-B.}\ \bibnamefont {Qiao}}, \ and\ \bibinfo
  {author} {\bibfnamefont {L.}~\bibnamefont {He}},\ }\href@noop {} {\bibfield
  {journal} {\bibinfo  {journal} {Nature Communications}\ }\textbf {\bibinfo
  {volume} {7}},\ \bibinfo {pages} {11760} (\bibinfo {year}
  {2016})}\BibitemShut {NoStop}%
\bibitem [{\citenamefont {Lee}\ \emph {et~al.}(2011)\citenamefont {Lee},
  \citenamefont {Riedl}, \citenamefont {Beringer}, \citenamefont {Neto},
  \citenamefont {von Klitzing}, \citenamefont {Starke},\ and\ \citenamefont
  {Smet}}]{lee2011quantum}%
  \BibitemOpen
  \bibfield  {author} {\bibinfo {author} {\bibfnamefont {D.~S.}\ \bibnamefont
  {Lee}}, \bibinfo {author} {\bibfnamefont {C.}~\bibnamefont {Riedl}}, \bibinfo
  {author} {\bibfnamefont {T.}~\bibnamefont {Beringer}}, \bibinfo {author}
  {\bibfnamefont {A.~C.}\ \bibnamefont {Neto}}, \bibinfo {author}
  {\bibfnamefont {K.}~\bibnamefont {von Klitzing}}, \bibinfo {author}
  {\bibfnamefont {U.}~\bibnamefont {Starke}}, \ and\ \bibinfo {author}
  {\bibfnamefont {J.~H.}\ \bibnamefont {Smet}},\ }\href@noop {} {\bibfield
  {journal} {\bibinfo  {journal} {Physical Review Letters}\ }\textbf {\bibinfo
  {volume} {107}},\ \bibinfo {pages} {216602} (\bibinfo {year}
  {2011})}\BibitemShut {NoStop}%
\bibitem [{\citenamefont {Sanchez-Yamagishi}\ \emph {et~al.}(2012)\citenamefont
  {Sanchez-Yamagishi}, \citenamefont {Taychatanapat}, \citenamefont {Watanabe},
  \citenamefont {Taniguchi}, \citenamefont {Yacoby},\ and\ \citenamefont
  {Jarillo-Herrero}}]{sanchez2012quantum}%
  \BibitemOpen
  \bibfield  {author} {\bibinfo {author} {\bibfnamefont {J.~D.}\ \bibnamefont
  {Sanchez-Yamagishi}}, \bibinfo {author} {\bibfnamefont {T.}~\bibnamefont
  {Taychatanapat}}, \bibinfo {author} {\bibfnamefont {K.}~\bibnamefont
  {Watanabe}}, \bibinfo {author} {\bibfnamefont {T.}~\bibnamefont {Taniguchi}},
  \bibinfo {author} {\bibfnamefont {A.}~\bibnamefont {Yacoby}}, \ and\ \bibinfo
  {author} {\bibfnamefont {P.}~\bibnamefont {Jarillo-Herrero}},\ }\href@noop {}
  {\bibfield  {journal} {\bibinfo  {journal} {Physical Review Letters}\
  }\textbf {\bibinfo {volume} {108}},\ \bibinfo {pages} {076601} (\bibinfo
  {year} {2012})}\BibitemShut {NoStop}%
\bibitem [{\citenamefont {Andrei}\ \emph {et~al.}(2012)\citenamefont {Andrei},
  \citenamefont {Li},\ and\ \citenamefont {Du}}]{andrei2012electronic}%
  \BibitemOpen
  \bibfield  {author} {\bibinfo {author} {\bibfnamefont {E.~Y.}\ \bibnamefont
  {Andrei}}, \bibinfo {author} {\bibfnamefont {G.}~\bibnamefont {Li}}, \ and\
  \bibinfo {author} {\bibfnamefont {X.}~\bibnamefont {Du}},\ }\href@noop {}
  {\bibfield  {journal} {\bibinfo  {journal} {Reports on Progress in Physics}\
  }\textbf {\bibinfo {volume} {75}},\ \bibinfo {pages} {056501} (\bibinfo
  {year} {2012})}\BibitemShut {NoStop}%
\bibitem [{\citenamefont {Cao}\ \emph {et~al.}(2016)\citenamefont {Cao},
  \citenamefont {Luo}, \citenamefont {Fatemi}, \citenamefont {Fang},
  \citenamefont {Sanchez-Yamagishi}, \citenamefont {Watanabe}, \citenamefont
  {Taniguchi}, \citenamefont {Kaxiras},\ and\ \citenamefont
  {Jarillo-Herrero}}]{cao2016superlattice}%
  \BibitemOpen
  \bibfield  {author} {\bibinfo {author} {\bibfnamefont {Y.}~\bibnamefont
  {Cao}}, \bibinfo {author} {\bibfnamefont {J.}~\bibnamefont {Luo}}, \bibinfo
  {author} {\bibfnamefont {V.}~\bibnamefont {Fatemi}}, \bibinfo {author}
  {\bibfnamefont {S.}~\bibnamefont {Fang}}, \bibinfo {author} {\bibfnamefont
  {J.}~\bibnamefont {Sanchez-Yamagishi}}, \bibinfo {author} {\bibfnamefont
  {K.}~\bibnamefont {Watanabe}}, \bibinfo {author} {\bibfnamefont
  {T.}~\bibnamefont {Taniguchi}}, \bibinfo {author} {\bibfnamefont
  {E.}~\bibnamefont {Kaxiras}}, \ and\ \bibinfo {author} {\bibfnamefont
  {P.}~\bibnamefont {Jarillo-Herrero}},\ }\href@noop {} {\bibfield  {journal}
  {\bibinfo  {journal} {Physical Review Letters}\ }\textbf {\bibinfo {volume}
  {117}},\ \bibinfo {pages} {116804} (\bibinfo {year} {2016})}\BibitemShut
  {NoStop}%
\bibitem [{\citenamefont {Kim}\ \emph {et~al.}(2017)\citenamefont {Kim},
  \citenamefont {DaSilva}, \citenamefont {Huang}, \citenamefont {Fallahazad},
  \citenamefont {Larentis}, \citenamefont {Taniguchi}, \citenamefont
  {Watanabe}, \citenamefont {LeRoy}, \citenamefont {MacDonald},\ and\
  \citenamefont {Tutuc}}]{kim2017tunable}%
  \BibitemOpen
  \bibfield  {author} {\bibinfo {author} {\bibfnamefont {K.}~\bibnamefont
  {Kim}}, \bibinfo {author} {\bibfnamefont {A.}~\bibnamefont {DaSilva}},
  \bibinfo {author} {\bibfnamefont {S.}~\bibnamefont {Huang}}, \bibinfo
  {author} {\bibfnamefont {B.}~\bibnamefont {Fallahazad}}, \bibinfo {author}
  {\bibfnamefont {S.}~\bibnamefont {Larentis}}, \bibinfo {author}
  {\bibfnamefont {T.}~\bibnamefont {Taniguchi}}, \bibinfo {author}
  {\bibfnamefont {K.}~\bibnamefont {Watanabe}}, \bibinfo {author}
  {\bibfnamefont {B.~J.}\ \bibnamefont {LeRoy}}, \bibinfo {author}
  {\bibfnamefont {A.~H.}\ \bibnamefont {MacDonald}}, \ and\ \bibinfo {author}
  {\bibfnamefont {E.}~\bibnamefont {Tutuc}},\ }\href@noop {} {\bibfield
  {journal} {\bibinfo  {journal} {Proceedings of the National Academy of
  Sciences USA}\ }\textbf {\bibinfo {volume} {114}},\ \bibinfo {pages} {3364}
  (\bibinfo {year} {2017})}\BibitemShut {NoStop}%
\bibitem [{\citenamefont {Li}\ \emph {et~al.}(2010{\natexlab{b}})\citenamefont
  {Li}, \citenamefont {Luican}, \citenamefont {Dos~Santos}, \citenamefont
  {Neto}, \citenamefont {Reina}, \citenamefont {Kong},\ and\ \citenamefont
  {Andrei}}]{li2010observation}%
  \BibitemOpen
  \bibfield  {author} {\bibinfo {author} {\bibfnamefont {G.}~\bibnamefont
  {Li}}, \bibinfo {author} {\bibfnamefont {A.}~\bibnamefont {Luican}}, \bibinfo
  {author} {\bibfnamefont {J.~L.}\ \bibnamefont {Dos~Santos}}, \bibinfo
  {author} {\bibfnamefont {A.~C.}\ \bibnamefont {Neto}}, \bibinfo {author}
  {\bibfnamefont {A.}~\bibnamefont {Reina}}, \bibinfo {author} {\bibfnamefont
  {J.}~\bibnamefont {Kong}}, \ and\ \bibinfo {author} {\bibfnamefont
  {E.}~\bibnamefont {Andrei}},\ }\href@noop {} {\bibfield  {journal} {\bibinfo
  {journal} {Nature Physics}\ }\textbf {\bibinfo {volume} {6}},\ \bibinfo
  {pages} {109} (\bibinfo {year} {2010}{\natexlab{b}})}\BibitemShut {NoStop}%
\bibitem [{\citenamefont {Brihuega}\ \emph {et~al.}(2012)\citenamefont
  {Brihuega}, \citenamefont {Mallet}, \citenamefont {Gonz{\'a}lez-Herrero},
  \citenamefont {de~Laissardi{\`e}re}, \citenamefont {Ugeda}, \citenamefont
  {Magaud}, \citenamefont {G{\'o}mez-Rodr{\'\i}guez}, \citenamefont
  {Yndur{\'a}in},\ and\ \citenamefont {Veuillen}}]{brihuega2012unraveling}%
  \BibitemOpen
  \bibfield  {author} {\bibinfo {author} {\bibfnamefont {I.}~\bibnamefont
  {Brihuega}}, \bibinfo {author} {\bibfnamefont {P.}~\bibnamefont {Mallet}},
  \bibinfo {author} {\bibfnamefont {H.}~\bibnamefont {Gonz{\'a}lez-Herrero}},
  \bibinfo {author} {\bibfnamefont {G.~T.}\ \bibnamefont
  {de~Laissardi{\`e}re}}, \bibinfo {author} {\bibfnamefont {M.}~\bibnamefont
  {Ugeda}}, \bibinfo {author} {\bibfnamefont {L.}~\bibnamefont {Magaud}},
  \bibinfo {author} {\bibfnamefont {J.}~\bibnamefont
  {G{\'o}mez-Rodr{\'\i}guez}}, \bibinfo {author} {\bibfnamefont
  {F.}~\bibnamefont {Yndur{\'a}in}}, \ and\ \bibinfo {author} {\bibfnamefont
  {J.-Y.}\ \bibnamefont {Veuillen}},\ }\href@noop {} {\bibfield  {journal}
  {\bibinfo  {journal} {Physical Review Letters}\ }\textbf {\bibinfo {volume}
  {109}},\ \bibinfo {pages} {196802} (\bibinfo {year} {2012})}\BibitemShut
  {NoStop}%
\bibitem [{\citenamefont {Yin}\ \emph {et~al.}(2015)\citenamefont {Yin},
  \citenamefont {Qiao}, \citenamefont {Zuo}, \citenamefont {Li},\ and\
  \citenamefont {He}}]{yin2015experimental}%
  \BibitemOpen
  \bibfield  {author} {\bibinfo {author} {\bibfnamefont {L.-J.}\ \bibnamefont
  {Yin}}, \bibinfo {author} {\bibfnamefont {J.-B.}\ \bibnamefont {Qiao}},
  \bibinfo {author} {\bibfnamefont {W.-J.}\ \bibnamefont {Zuo}}, \bibinfo
  {author} {\bibfnamefont {W.-T.}\ \bibnamefont {Li}}, \ and\ \bibinfo {author}
  {\bibfnamefont {L.}~\bibnamefont {He}},\ }\href@noop {} {\bibfield  {journal}
  {\bibinfo  {journal} {Physical Review B}\ }\textbf {\bibinfo {volume} {92}},\
  \bibinfo {pages} {081406} (\bibinfo {year} {2015})}\BibitemShut {NoStop}%
\bibitem [{\citenamefont {Wong}\ \emph {et~al.}(2015)\citenamefont {Wong},
  \citenamefont {Wang}, \citenamefont {Jung}, \citenamefont {Pezzini},
  \citenamefont {DaSilva}, \citenamefont {Tsai}, \citenamefont {Jung},
  \citenamefont {Khajeh}, \citenamefont {Kim}, \citenamefont {Lee} \emph
  {et~al.}}]{wong2015local}%
  \BibitemOpen
  \bibfield  {author} {\bibinfo {author} {\bibfnamefont {D.}~\bibnamefont
  {Wong}}, \bibinfo {author} {\bibfnamefont {Y.}~\bibnamefont {Wang}}, \bibinfo
  {author} {\bibfnamefont {J.}~\bibnamefont {Jung}}, \bibinfo {author}
  {\bibfnamefont {S.}~\bibnamefont {Pezzini}}, \bibinfo {author} {\bibfnamefont
  {A.~M.}\ \bibnamefont {DaSilva}}, \bibinfo {author} {\bibfnamefont {H.-Z.}\
  \bibnamefont {Tsai}}, \bibinfo {author} {\bibfnamefont {H.~S.}\ \bibnamefont
  {Jung}}, \bibinfo {author} {\bibfnamefont {R.}~\bibnamefont {Khajeh}},
  \bibinfo {author} {\bibfnamefont {Y.}~\bibnamefont {Kim}}, \bibinfo {author}
  {\bibfnamefont {J.}~\bibnamefont {Lee}},  \emph {et~al.},\ }\href@noop {}
  {\bibfield  {journal} {\bibinfo  {journal} {Physical Review B}\ }\textbf
  {\bibinfo {volume} {92}},\ \bibinfo {pages} {155409} (\bibinfo {year}
  {2015})}\BibitemShut {NoStop}%
\bibitem [{\citenamefont {Kim}\ \emph {et~al.}(2016)\citenamefont {Kim},
  \citenamefont {Yankowitz}, \citenamefont {Fallahazad}, \citenamefont {Kang},
  \citenamefont {Movva}, \citenamefont {Huang}, \citenamefont {Larentis},
  \citenamefont {Corbet}, \citenamefont {Taniguchi}, \citenamefont {Watanabe}
  \emph {et~al.}}]{kim2016van}%
  \BibitemOpen
  \bibfield  {author} {\bibinfo {author} {\bibfnamefont {K.}~\bibnamefont
  {Kim}}, \bibinfo {author} {\bibfnamefont {M.}~\bibnamefont {Yankowitz}},
  \bibinfo {author} {\bibfnamefont {B.}~\bibnamefont {Fallahazad}}, \bibinfo
  {author} {\bibfnamefont {S.}~\bibnamefont {Kang}}, \bibinfo {author}
  {\bibfnamefont {H.~C.}\ \bibnamefont {Movva}}, \bibinfo {author}
  {\bibfnamefont {S.}~\bibnamefont {Huang}}, \bibinfo {author} {\bibfnamefont
  {S.}~\bibnamefont {Larentis}}, \bibinfo {author} {\bibfnamefont {C.~M.}\
  \bibnamefont {Corbet}}, \bibinfo {author} {\bibfnamefont {T.}~\bibnamefont
  {Taniguchi}}, \bibinfo {author} {\bibfnamefont {K.}~\bibnamefont {Watanabe}},
   \emph {et~al.},\ }\href@noop {} {\bibfield  {journal} {\bibinfo  {journal}
  {Nano Letters}\ }\textbf {\bibinfo {volume} {16}},\ \bibinfo {pages} {1989}
  (\bibinfo {year} {2016})}\BibitemShut {NoStop}%
\bibitem [{SI()}]{SI}%
  \BibitemOpen
  \href@noop {} {\ }\bibinfo {note} {See Supplemental Material at [URL will be
  inserted by publisher] for extra experimental data, data analysis and
  theoretical calculation, which includes Refs. [33-39].}\BibitemShut {Stop}%
\bibitem [{\citenamefont {{Efimkin}}\ and\ \citenamefont
  {{MacDonald}}()}]{EfimkinNetwork}%
  \BibitemOpen
  \bibfield  {author} {\bibinfo {author} {\bibfnamefont {D.~K.}\ \bibnamefont
  {{Efimkin}}}\ and\ \bibinfo {author} {\bibfnamefont {A.~H.}\ \bibnamefont
  {{MacDonald}}},\ }\href@noop {} {\ }\Eprint {http://arxiv.org/abs/1803.06404
  (2018)} {arXiv:1803.06404 (2018)} \BibitemShut {NoStop}%
\bibitem [{\citenamefont {Bistritzer}\ and\ \citenamefont
  {MacDonald}(2011)}]{bistritzer2011moire}%
  \BibitemOpen
  \bibfield  {author} {\bibinfo {author} {\bibfnamefont {R.}~\bibnamefont
  {Bistritzer}}\ and\ \bibinfo {author} {\bibfnamefont {A.~H.}\ \bibnamefont
  {MacDonald}},\ }\href@noop {} {\bibfield  {journal} {\bibinfo  {journal}
  {Proceedings of the National Academy of Sciences}\ }\textbf {\bibinfo
  {volume} {108}},\ \bibinfo {pages} {12233} (\bibinfo {year}
  {2011})}\BibitemShut {NoStop}%
\bibitem [{\citenamefont {Trambly~de Laissardiere}\ \emph
  {et~al.}(2010)\citenamefont {Trambly~de Laissardiere}, \citenamefont
  {Mayou},\ and\ \citenamefont {Magaud}}]{trambly2010localization}%
  \BibitemOpen
  \bibfield  {author} {\bibinfo {author} {\bibfnamefont {G.}~\bibnamefont
  {Trambly~de Laissardiere}}, \bibinfo {author} {\bibfnamefont
  {D.}~\bibnamefont {Mayou}}, \ and\ \bibinfo {author} {\bibfnamefont
  {L.}~\bibnamefont {Magaud}},\ }\href@noop {} {\bibfield  {journal} {\bibinfo
  {journal} {Nano Letters}\ }\textbf {\bibinfo {volume} {10}},\ \bibinfo
  {pages} {804} (\bibinfo {year} {2010})}\BibitemShut {NoStop}%
\bibitem [{\citenamefont {Yankowitz}\ \emph {et~al.}(2014)\citenamefont
  {Yankowitz}, \citenamefont {Wang}, \citenamefont {Li}, \citenamefont
  {Birdwell}, \citenamefont {Chen}, \citenamefont {Watanabe}, \citenamefont
  {Taniguchi}, \citenamefont {Quek}, \citenamefont {Jarillo-Herrero},\ and\
  \citenamefont {LeRoy}}]{yankowitz2014band}%
  \BibitemOpen
  \bibfield  {author} {\bibinfo {author} {\bibfnamefont {M.}~\bibnamefont
  {Yankowitz}}, \bibinfo {author} {\bibfnamefont {J.~I.-J.}\ \bibnamefont
  {Wang}}, \bibinfo {author} {\bibfnamefont {S.}~\bibnamefont {Li}}, \bibinfo
  {author} {\bibfnamefont {A.~G.}\ \bibnamefont {Birdwell}}, \bibinfo {author}
  {\bibfnamefont {Y.-A.}\ \bibnamefont {Chen}}, \bibinfo {author}
  {\bibfnamefont {K.}~\bibnamefont {Watanabe}}, \bibinfo {author}
  {\bibfnamefont {T.}~\bibnamefont {Taniguchi}}, \bibinfo {author}
  {\bibfnamefont {S.~Y.}\ \bibnamefont {Quek}}, \bibinfo {author}
  {\bibfnamefont {P.}~\bibnamefont {Jarillo-Herrero}}, \ and\ \bibinfo {author}
  {\bibfnamefont {B.~J.}\ \bibnamefont {LeRoy}},\ }\href@noop {} {\bibfield
  {journal} {\bibinfo  {journal} {APL Materials}\ }\textbf {\bibinfo {volume}
  {2}},\ \bibinfo {pages} {092503} (\bibinfo {year} {2014})}\BibitemShut
  {NoStop}%
\bibitem [{\citenamefont {Castro}\ \emph {et~al.}(2007)\citenamefont {Castro},
  \citenamefont {Novoselov}, \citenamefont {Morozov}, \citenamefont {Peres},
  \citenamefont {Dos~Santos}, \citenamefont {Nilsson}, \citenamefont {Guinea},
  \citenamefont {Geim},\ and\ \citenamefont {Neto}}]{castro2007biased}%
  \BibitemOpen
  \bibfield  {author} {\bibinfo {author} {\bibfnamefont {E.~V.}\ \bibnamefont
  {Castro}}, \bibinfo {author} {\bibfnamefont {K.}~\bibnamefont {Novoselov}},
  \bibinfo {author} {\bibfnamefont {S.}~\bibnamefont {Morozov}}, \bibinfo
  {author} {\bibfnamefont {N.}~\bibnamefont {Peres}}, \bibinfo {author}
  {\bibfnamefont {J.~L.}\ \bibnamefont {Dos~Santos}}, \bibinfo {author}
  {\bibfnamefont {J.}~\bibnamefont {Nilsson}}, \bibinfo {author} {\bibfnamefont
  {F.}~\bibnamefont {Guinea}}, \bibinfo {author} {\bibfnamefont
  {A.}~\bibnamefont {Geim}}, \ and\ \bibinfo {author} {\bibfnamefont {A.~C.}\
  \bibnamefont {Neto}},\ }\href@noop {} {\bibfield  {journal} {\bibinfo
  {journal} {Physical review letters}\ }\textbf {\bibinfo {volume} {99}},\
  \bibinfo {pages} {216802} (\bibinfo {year} {2007})}\BibitemShut {NoStop}%
\bibitem [{\citenamefont {Kuzmenko}\ \emph {et~al.}(2009)\citenamefont
  {Kuzmenko}, \citenamefont {Van~Heumen}, \citenamefont {Van Der~Marel},
  \citenamefont {Lerch}, \citenamefont {Blake}, \citenamefont {Novoselov},\
  and\ \citenamefont {Geim}}]{kuzmenko2009infrared}%
  \BibitemOpen
  \bibfield  {author} {\bibinfo {author} {\bibfnamefont {A.}~\bibnamefont
  {Kuzmenko}}, \bibinfo {author} {\bibfnamefont {E.}~\bibnamefont
  {Van~Heumen}}, \bibinfo {author} {\bibfnamefont {D.}~\bibnamefont {Van
  Der~Marel}}, \bibinfo {author} {\bibfnamefont {P.}~\bibnamefont {Lerch}},
  \bibinfo {author} {\bibfnamefont {P.}~\bibnamefont {Blake}}, \bibinfo
  {author} {\bibfnamefont {K.}~\bibnamefont {Novoselov}}, \ and\ \bibinfo
  {author} {\bibfnamefont {A.}~\bibnamefont {Geim}},\ }\href@noop {} {\bibfield
   {journal} {\bibinfo  {journal} {Physical Review B}\ }\textbf {\bibinfo
  {volume} {79}},\ \bibinfo {pages} {115441} (\bibinfo {year}
  {2009})}\BibitemShut {NoStop}%
\bibitem [{\citenamefont {Li}\ \emph {et~al.}(2009)\citenamefont {Li},
  \citenamefont {Henriksen}, \citenamefont {Jiang}, \citenamefont {Hao},
  \citenamefont {Martin}, \citenamefont {Kim}, \citenamefont {Stormer},\ and\
  \citenamefont {Basov}}]{li2009band}%
  \BibitemOpen
  \bibfield  {author} {\bibinfo {author} {\bibfnamefont {Z.}~\bibnamefont
  {Li}}, \bibinfo {author} {\bibfnamefont {E.}~\bibnamefont {Henriksen}},
  \bibinfo {author} {\bibfnamefont {Z.}~\bibnamefont {Jiang}}, \bibinfo
  {author} {\bibfnamefont {Z.}~\bibnamefont {Hao}}, \bibinfo {author}
  {\bibfnamefont {M.~C.}\ \bibnamefont {Martin}}, \bibinfo {author}
  {\bibfnamefont {P.}~\bibnamefont {Kim}}, \bibinfo {author} {\bibfnamefont
  {H.}~\bibnamefont {Stormer}}, \ and\ \bibinfo {author} {\bibfnamefont
  {D.~N.}\ \bibnamefont {Basov}},\ }\href@noop {} {\bibfield  {journal}
  {\bibinfo  {journal} {Physical Review Letters}\ }\textbf {\bibinfo {volume}
  {102}},\ \bibinfo {pages} {037403} (\bibinfo {year} {2009})}\BibitemShut
  {NoStop}%
\bibitem [{\citenamefont {Zhang}\ \emph {et~al.}(2008)\citenamefont {Zhang},
  \citenamefont {Li}, \citenamefont {Basov}, \citenamefont {Fogler},
  \citenamefont {Hao},\ and\ \citenamefont {Martin}}]{zhang2008determination}%
  \BibitemOpen
  \bibfield  {author} {\bibinfo {author} {\bibfnamefont {L.}~\bibnamefont
  {Zhang}}, \bibinfo {author} {\bibfnamefont {Z.}~\bibnamefont {Li}}, \bibinfo
  {author} {\bibfnamefont {D.~N.}\ \bibnamefont {Basov}}, \bibinfo {author}
  {\bibfnamefont {M.}~\bibnamefont {Fogler}}, \bibinfo {author} {\bibfnamefont
  {Z.}~\bibnamefont {Hao}}, \ and\ \bibinfo {author} {\bibfnamefont {M.~C.}\
  \bibnamefont {Martin}},\ }\href@noop {} {\bibfield  {journal} {\bibinfo
  {journal} {Physical Review B}\ }\textbf {\bibinfo {volume} {78}},\ \bibinfo
  {pages} {235408} (\bibinfo {year} {2008})}\BibitemShut {NoStop}%
\bibitem [{\citenamefont {Bl\"ochl}\ \emph {et~al.}(1994)\citenamefont
  {Bl\"ochl}, \citenamefont {Jepsen},\ and\ \citenamefont
  {Andersen}}]{blochl1994tetrahedron}%
  \BibitemOpen
  \bibfield  {author} {\bibinfo {author} {\bibfnamefont {P.~E.}\ \bibnamefont
  {Bl\"ochl}}, \bibinfo {author} {\bibfnamefont {O.}~\bibnamefont {Jepsen}}, \
  and\ \bibinfo {author} {\bibfnamefont {O.~K.}\ \bibnamefont {Andersen}},\
  }\href {\doibase 10.1103/PhysRevB.49.16223} {\bibfield  {journal} {\bibinfo
  {journal} {Phys. Rev. B}\ }\textbf {\bibinfo {volume} {49}},\ \bibinfo
  {pages} {16223} (\bibinfo {year} {1994})}\BibitemShut {NoStop}%
\end{thebibliography}%

\end{document}